\documentclass[%
 reprint,
superscriptaddress,
 amsmath,
 amssymb,
 aps,
 prx,
 showpacs, 
 floatfix,
]{revtex4-2}

\usepackage{graphicx}
\usepackage{dcolumn}
\usepackage{bm}

\usepackage{hyperref}
\hypersetup{
  colorlinks   = true, 
  urlcolor     = blue, 
  linkcolor    = blue, 
  citecolor   = blue 
}

\begin{document}

\title{Planckian Diffusion: The Ghost of Anderson Localization }

\author{Yubo~Zhang} 
\affiliation{School of Physics, Peking University, Beijing 100871, China}

\author{Anton M.~Graf}
\affiliation{Harvard John A. Paulson School of Engineering and Applied Sciences,
Harvard, Cambridge, Massachusetts 02138, USA}
\affiliation{Department of Chemistry and Chemical Biology, Harvard University, Cambridge,
Massachusetts 02138, USA}

\author{Alhun~Aydin}
\affiliation{Department of Physics, Harvard University, Cambridge, Massachusetts 02138, USA}
\affiliation{Faculty of Engineering and Natural Sciences, Sabanci University, 34956 Tuzla, Istanbul, Turkey}

\author{Joonas~Keski-Rahkonen}
\affiliation{Department of Chemistry and Chemical Biology, Harvard University, Cambridge,
Massachusetts 02138, USA}
\affiliation{Department of Physics, Harvard University, Cambridge, Massachusetts 02138, USA}

\author{Eric J.~Heller} 
\affiliation{Department of Chemistry and Chemical Biology, Harvard University, Cambridge,
Massachusetts 02138, USA}
\affiliation{Department of Physics, Harvard University, Cambridge, Massachusetts 02138, USA}

\date{\today}

\begin{abstract}
We find that Anderson localization  ceases to exist when a random medium begins to move, but another type of fundamental quantum effect, Planckian diffusion $D = \alpha\hbar/m$, rises to replace it, with $\alpha $ of order of unity.   Planckian diffusion supercedes the Planckian speed limit $\tau= \alpha \hbar/k_B T,$ as it not only implies this relation in thermal systems but also applies more generally without requiring thermal equilibrium. Here we model a dynamic disordered system with thousands of itinerant impurities, having random initial positions and velocities. By incrementally increasing their speed from zero, we observe a transition from Anderson localization to Planckian diffusion, with  $\alpha$  falling within the range of $0.5$ to $2$.  Furthermore, we relate the breakdown of Anderson localization to three additional, distinctly different confirming cases that also exhibit Planckian diffusion $D\sim \hbar/m$, including one experiment on solid hydrogen. Our finding suggests that Planckian diffusion in dynamic disordered systems is as universal as Anderson localization in static disordered systems, which may shed light on quantum transport studies.

\end{abstract}

\maketitle

\section{Introduction}






 In 1958, P.W. Anderson proposed that quantum wavefunction localization is a  universal phenomenon in which waves stop diffusing in static disordered systems~\cite{anderson1958absence, kramer1993localization}. It can be understood as a wave interference phenomenon~\cite{lagendijk2009fifty}. In a perfectly ordered medium, particles can move freely, forming extended wave functions. However, in a disordered medium, particles scatter multiple times off impurities. Each scattering event alters the phase of the particle's wave function. At a certain level of disorder, when these scattered waves interfere strongly and destructively, they can prevent the wave from propagating further, effectively ``trapping" the particle in a localized region. Over the past two decades, Anderson localization has been observed across a variety of platforms, including ultracold atomic gases~\cite{roati2008anderson}, photonic lattices~\cite{schwartz2007transport}, and acoustic systems~\cite{hu2008localization}, encompassing one-dimensional~\cite{billy2008direct}, two-dimensional~\cite{white_observation_2020} and three-dimensional geometries~\cite{kondov2011three, jendrzejewski2012three}. 


 However, the phase violence introduced by a motion of the random potential scrambles the quantum interference required for the Anderson localization. As a result, diffusion is the natural outcome in dynamic disordered systems.  On the other hand, the sporadic rise and fall of the Anderson localization can result in what has been called transient localization. This has been of interest in the context of crystalline organic semiconductors and halide perovskites~\cite{afmTL}. Furthermore, some of the mysterious features of strange metals have recently been shown to emerge from the fugitive Anderson localization caused by a morphing deformation potential landscape under thermal lattice vibrations, within the standard Fr{\"o}hlich model~\cite{heller22,zimmermann_rise_2024,w,Aydin24}.

In this study, we introduce time dependence to a random potential, thereby disrupting Anderson localization and reinstating diffusion. Provided that the random potential evolves at a sufficiently rapid rate, our key finding is the emergence of a universal diffusion characterized by the Planckian diffusion coefficient  $D = \alpha \hbar/m$, where $\alpha $ is a dimensionless parameter of order unity and $m$ is the effective mass of the diffusing particle. We claim this behavior ``universal'' in the sense that the diffusion coefficient consistently approximates $\hbar/m$ for a wide parameter space. This universality is maintained despite variations in the medium's motion speed, the coupling strength, and the system's temperature within a thermal model framework.

\section{Model}

We investigate a two-dimensional system with thousands of impurities modeled by potential bumps, each represented by a Bessel function multiplied by an exponentially decaying factor which makes them localized. The peak potential height of each bump is on the order of $1$~eV  and the typical size of each bump center is in the order of $1$~nm, with tails extending to around $10$~nm. Their initial positions and velocity directions are generated randomly. Speed is set to be an adjustable constant; they move classically without interacting with each other.


Within the time-dependent maelstrom described above, we launch a Gaussian wavepacket representing a charge carrier with initial width $10$ nm and zero average momentum. We find the subsequent carrier dynamics utilizing the split-operator method, computing the mean square diffusion radius as a function of time~\cite{aydin_quantum_2024}:
$$\Delta R(t) = \int \psi^*(\textbf{r},t)[\textbf{r}(t)-\bar{\textbf{r}}(t)]^2 \psi(\textbf{r},t)\, \textrm{d}\mathbf{r}$$
where $\psi(\textbf{r},t)$ is the wavefunction in real space at time t and position $\textbf{r}$, and $\bar{\textbf{r}}(t)$ is the average position traveled at time t. In our simulation, $\bar{\textbf{r}}(t)$ is always near zero, due to the randomness of impurities and zero initial central momentum of the wave packet. We also checked that a non-zero (but relatively small) initial central momentum does not affect the MSD result, however, to make the most use of the window size and delay it from reaching the boundary, we set it to be zero. We evaluate the diffusion coefficient \cite{aydin_quantum_2024}
\[D =  \lim_{t\to\infty} \frac{1}{2d} \frac{\partial \Delta R(t)}{\partial t}\] 
where $d=2$ is the dimension of the system. The long-time limit ensures that initial conditions are forgotten, since a Gaussian wave packet includes various components of momentum.

\section{Results}

\subsection{Anderson localization and adiabatic limit}

We initially set the impurity velocities to zero. As expected, in such a static disordered system, Anderson localization emerges. As shown in Figure~\ref{statdyn} (a) and (b), both the strongly confined wavefunction itself and the almost horizontal behavior of mean square distance against time indicate absence of diffusion, confirming that we have Anderson localization in our model.

The initial diffusion arises from the Gaussian wave packet we launched, which is not an eigenstate and therefore undergoes spreading. As time evolves, the mean square distance curve progressively flattens, signifying a reduction in diffusion and ultimately leading to an Anderson localized state.

Stronger potentials lead to shorter localization lengths and reduce the time required for the wave function to reach equilibrium: zero diffusion. While weaker random potentials are sufficient to localize the wave function over extended times and larger length scales, we employed strong potentials with peak values on the order of 1 eV to optimize the use of our computational window size and to reduce computational time.

\subsection{Planckian diffusion domain and beyond}

\begin{figure}[h]
    \centering
    \includegraphics[width=1\linewidth]{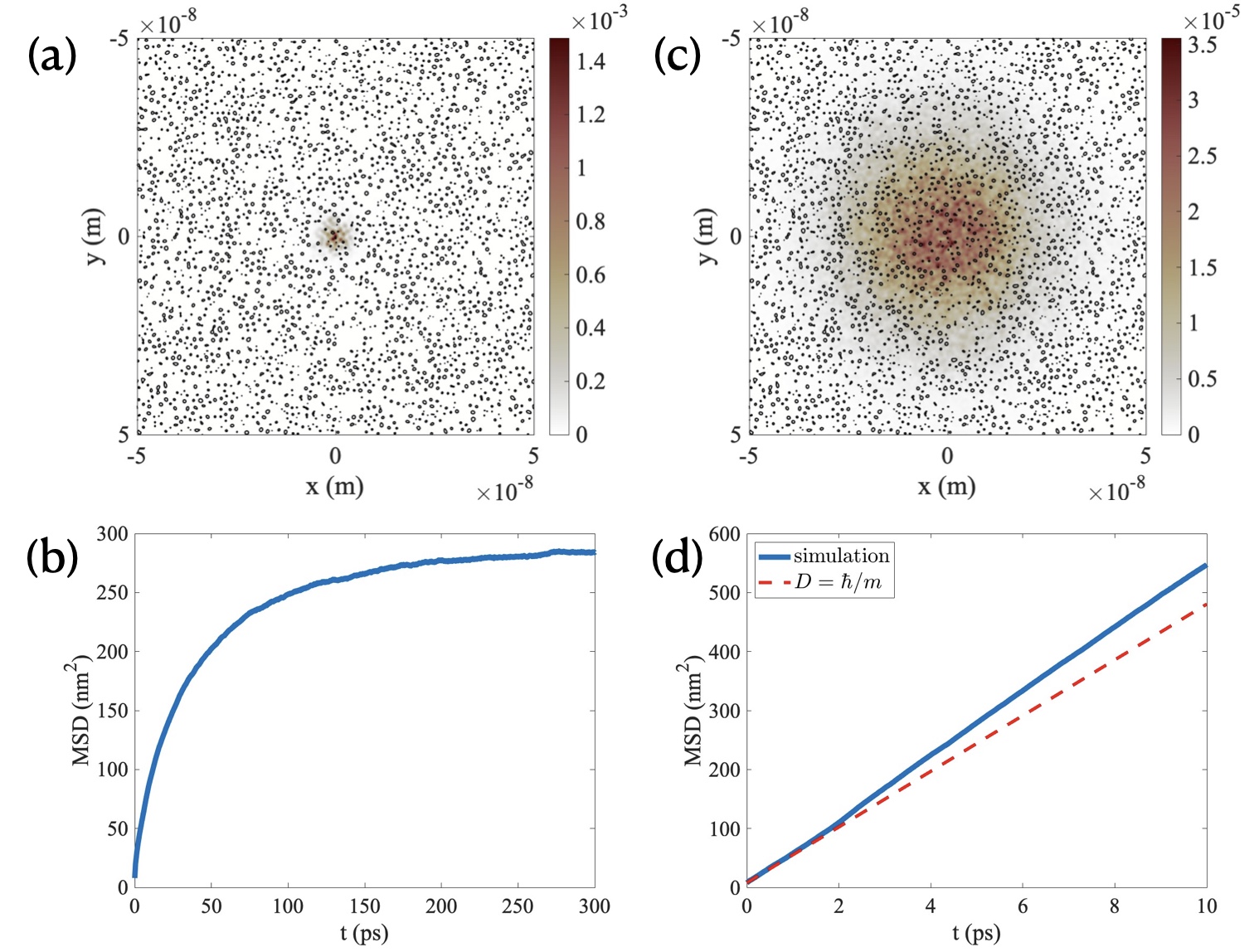}
    \caption{Static impurities versus dynamical impurities. Randomly distributed frozen scatterers leads to localization of wave function, verified by the strongly confined picture of probability density in real space during long time evolution (a), and also by nearly horizontal line of mean square distance after 200 ps (b). Randomly moving scatterers leads to diffusive behavior of wavefunction in Planckian rate (c), verified by the linear behavior of mean square distance against time, which has a slope near $\hbar/m$ (dashed line),  (d).}
    \label{statdyn}
\end{figure}

For the right hand column in Figure~\ref{statdyn}, we give the impurities  a constant speed but random directions ($v=2000$ m/s corresponding roughly to a sound velocity), with all the other potential parameters kept the same, near those for an electron in an optimally doped cuprate.  The wavefunction diffuses with a constant rate, near $\hbar/m$.

\begin{figure}[h]
    \centering
    \includegraphics[width=1\linewidth]{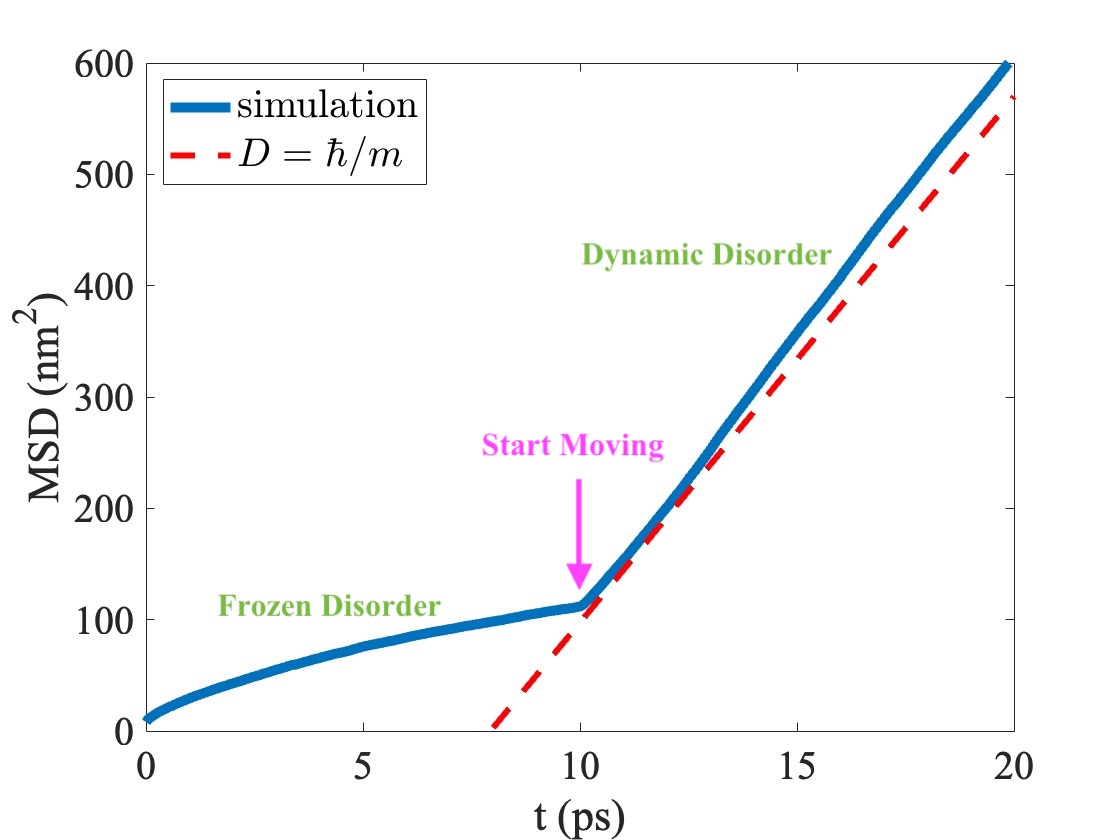}
    \caption{Activation of disorder dynamics at 10 ps. Prior to 10 ps, the diffusion coefficient gradually saturates to zero, indicating Anderson localization in a static disorder environment. At 10 ps, the disorder potential starts to move and the diffusion coefficient $D$ abruptly  jumps to $\hbar/m$,  the Planckian diffusion regime in a dynamic disorder environment.}
    \label{transition}
\end{figure}

It is essential to understand the transition in wavefunction behavior as the system shifts from a static to a moving medium. When the impurity potential remains at rest, the wavefunction gradually saturates into a localized state, with the diffusion coefficient steadily decreasing over time. At 10 ps, the motion of the medium is introduced, marking a dramatic transition. Once the medium begins to move, the wavefunction rapidly shifts to diffusing at a constant speed, approximately $\hbar/m$, as illustrated in Figure~\ref{transition}.

\begin{figure}[h]
    \centering
    \includegraphics[width=1\linewidth]{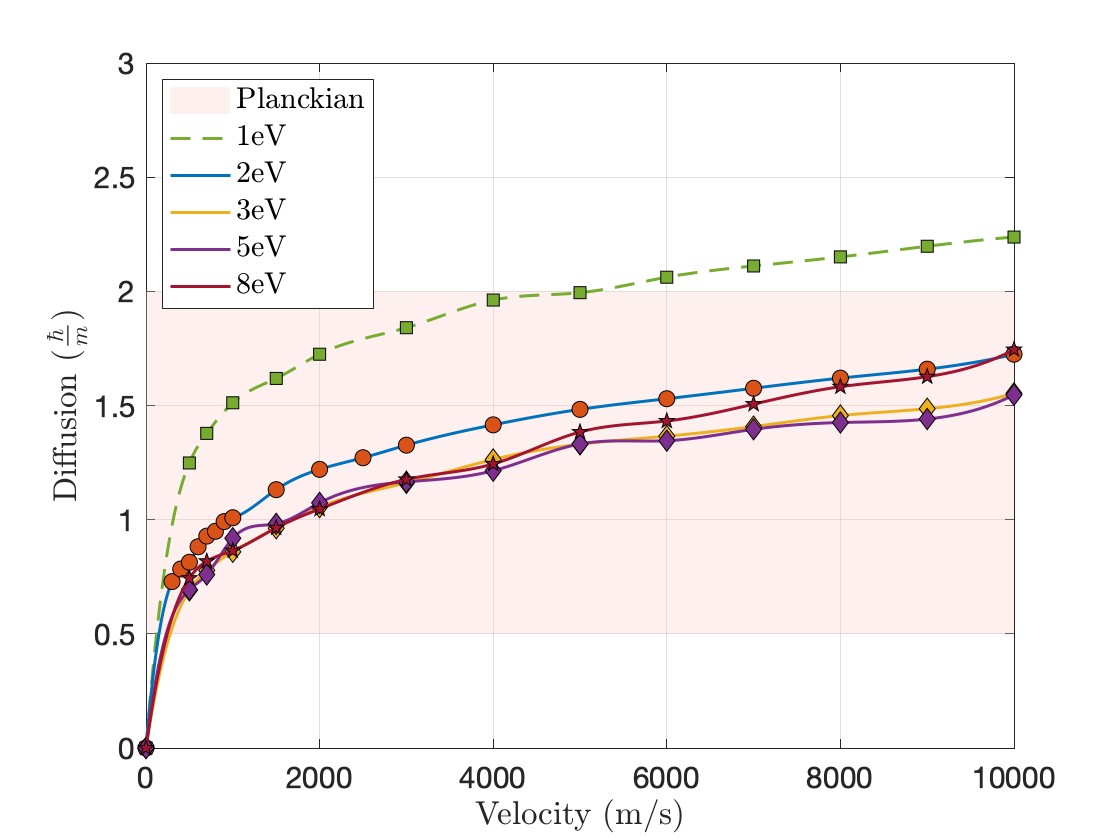}
    \caption{Dependence of diffusion on impurity velocity and impurity potential. It  immediately rises from the Anderson localization phase,    quickly transitions through adiabatic phase, and   rests in the Planckian diffusion regime for a wide range of velocities, showing robustness and implying that Planckian diffusion is natural outcome when Anderson localization is broken down by moving impurities.}
    \label{velocity}
\end{figure}

To test the robustness of this finding, we  found diffusion coefficients for different impurity velocities, potential heights, and  distributions of velocities. When varying potential heights, we maintain the total cross section, which is proportional to the product of potential height and impurity densities. 
We see from Figure~\ref{velocity} that as the fixed impurity velocity increases, the diffusion coefficient quickly rises from zero to the Planckian regime, defined as $0.5 \sim 2 \hbar/m$, highlighted by the pink box, supporting the claim that Planckian diffusion results when Anderson localization is broken by moving impurities.

Starting from the zero diffusion and zero impurity velocity of Anderson localization regime, with very slowly moving impurities (slow relative to the carrier velocity) the system enters a nonuniversal and semi-adiabatic regime with diffusion  $D \ll \hbar/m$ depending on impurity velocity. The charge carrier wavefunction slowly makes a transition from one equilibrium state to another with respect to spatially proximate Anderson localized states, diffusing at a rate proportional to rate of change of the potential. It cannot be adiabatic with respect to the very many, exponentially weakly coupled distant states at any finite velocity however, thus we call this regime semi-adiabatic. If it were not so narrow a region as a function of impurity speed, the semi-adiabatic regime might be more interesting in its own right. No exact data points are plotted in the adiabatic  regime in Figure~\ref{velocity}, because for such very  slow movement we cannot accurately  gauge diffusion. The semi-adiabatic regime ends when  hops to spatially adjacent Anderson states are not followed adiabatically.

\begin{figure}[h]
    \centering
    \includegraphics[width=1\linewidth]{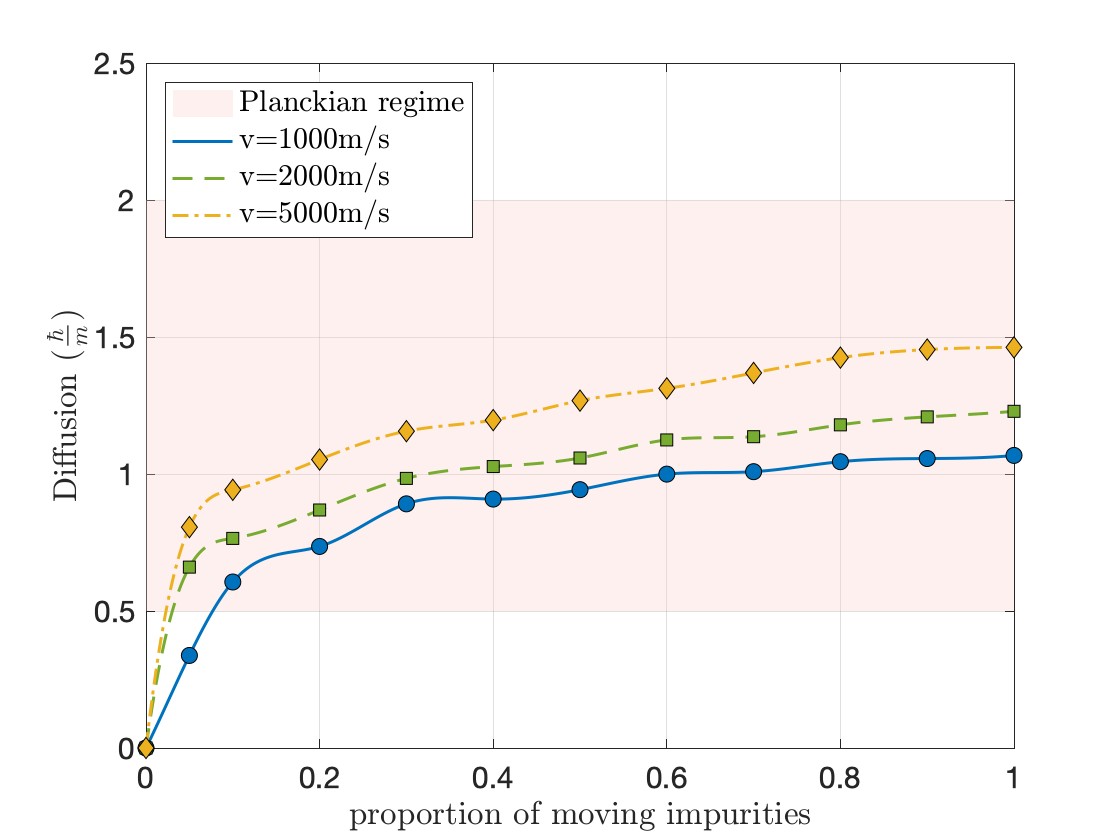}
    \caption{Dependence of diffusion on proportion of moving impurities and velocity. $10\%$ of moving impurities is enough to break Anderson localization caused by other $90\%$ fixed impurities. Planckian regime is highlighted with a red box. This could provide more general explanations for experimental results.}
    \label{proportion}
\end{figure}

The narrow semi-adiabatic regime is followed by a broad ghost regime featuring Planckian diffusion,  defined as $D \sim 0.5 \leftrightarrow  2\hbar/m$. Some indication of how broad the ghost regime is can be seen in figures~\ref{velocity}, \ref{proportion}, and~\ref{maxwell}.

Finally, there is the classical-like regime, defined as $D \gg 2\hbar/m$, realized when impurities velocities are extremely high, as seen in the normally visible world around us. When the impurities are very fast, then within the frame of an impurity, collisions with the carrier are nearly classical, with interference (quantum) effects becoming ignorable.                                                       



Not all impurities need to be in motion to break Anderson localization. We find that if $10\%$ are moving, it is enough to cause diffusive behavior at the Planckian rate. The results are shown in Figure~\ref{proportion}. As the proportion of moving impurities increases, diffusion coefficient quickly jumps up from zero to around $\hbar/m$, and stays in that Planckian regime. The higher the percentage of movers, the higher is the diffusion, but the diffusion remains near Planckian. It is quite evident that a moving medium prevents a strict, localizing coherence to be established, but the details of this disruption may one day illuminate the intricaces of Anderson localization.  


In Anderson localization, the interference of multiply scattered waves leads to the localization of electrons in a disordered medium. By  introducing some randomly moving impurities  it is possible to destroy such phase relations on a timescale  short enough to break Anderson localization, before it can be set up. The time required to set localization up will become a key target of opportunity in future studies of ghost Planckian diffusion.

In particular, our simulations also reveal that in the case of frozen disorder, a longer localization length—resulting from a weaker random potential—requires a lower velocity to disrupt localization and achieve diffusion at the Planckian rate. A shorter localization length corresponds to higher speed of motion of impurities, to reach Planckian regime. This outcome naturally aligns with the role of impurities in destroying coherence of wavefunction.

\subsection{Boltzmannian impurities}

\begin{figure}[h]
    \centering
    \includegraphics[width=1\linewidth]{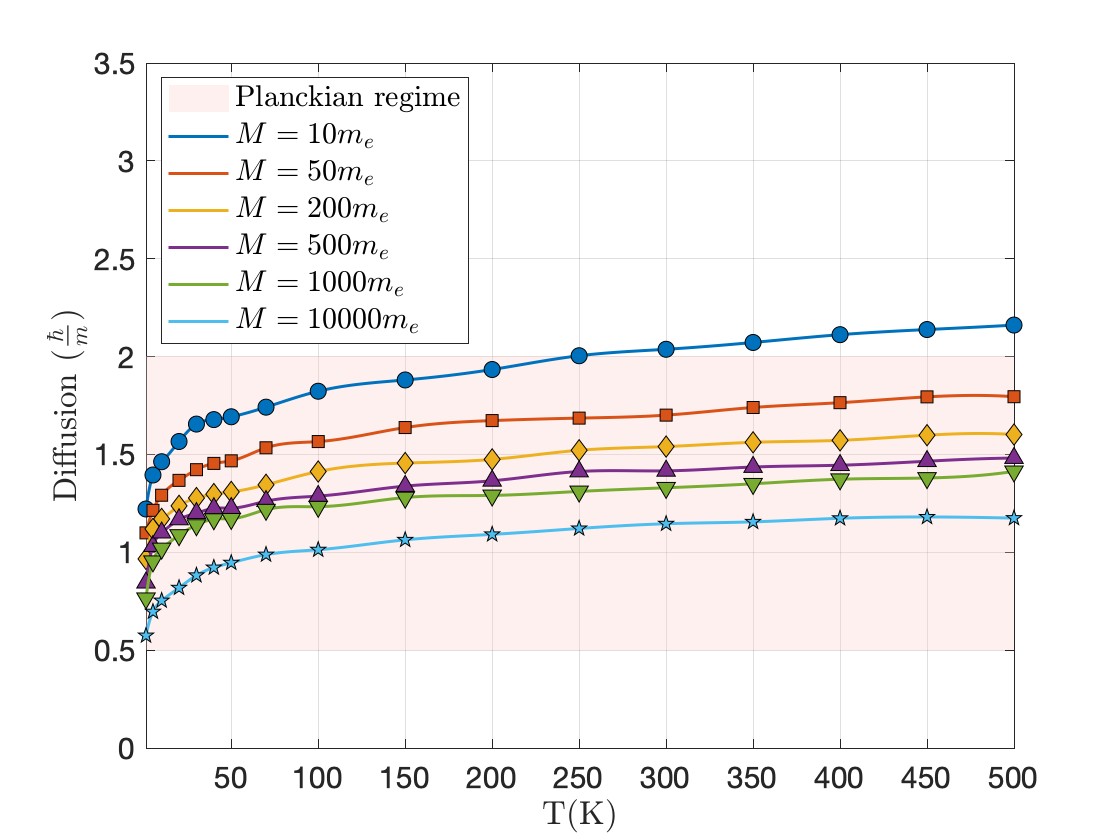}
    \caption{Planckian regime is highlighted with pink stripe. Using a Maxwell velocity distribution of the impurities, at various temperatures, and various masses, the Planckian diffusive behavior is robust. As temperature varies from 1K to 500K, and impurity masses vary from 10 to 10000 times electron mass, the diffusion constant always stays around $\hbar/m$, indicating universality. }
    \label{maxwell}
\end{figure}

We next adjust our model, making the impurities follow a Maxwell-Boltzmann  isotropic distribution of velocities.  This introduces a  temperature dependence. We calculate diffusion coefficients at various temperatures and masses of impurities, shown in Figure~\ref{maxwell}. Diffusion is robust against any changes in temperature from $1$ K to $500$~K and mass from $10m_e$ to $10000m_e$, further supporting the universality of Planckian diffusion in dynamic disordered systems. 

This result is consistent with the earlier model, where we saw that  diffusion is robust to impurity velocity as long as it is not too slow. Even for the $1$ K case, although the most probable velocity  is around $200$ m/s, very small,   the tail of Maxwell distribution,   includes $20\%$ of the particles that possess velocities higher than $500$ m/s.  This is more than  is enough to lead to Planckian diffusion. 

In Figure~\ref{maxwell}, the Bessel-like potential for each impurity is $2$ eV in the center height, but the center is not the most important part of the defect.  Most of the scattering comes from the ``skirts'' of height $0.3$ eV surrounding the peak. In the simulation, the impurity density is set to $0.4 \;\text{nm}^{-2}$, which is 2 order of magnitude higher than normal materials~\cite{chung2021ultra} but can be found in Si:P $\delta$-doped layers. This system typically exhibit impurity densities in the range of $10^{13} \sim 10^{14} \;\text{cm}^{-2}$~\cite{hwang2013electronic}. These realistic parameters combined with the robustness of diffusion against changes of impurity mass and temperature, provide us the confidence to claim that Planckian diffusion in dynamic disordered systems is as universal as Anderson localization in static disordered systems.

We connect our diffusion coefficient with a scattering rate, using an Einstein relation~\cite{wetzelaer_validity_2011}, 
and we find 
$$\frac{1}{\tau} = \frac{k_B T}{m D} $$
where $\tau$ is the scattering time defined in Drude model.  Then  resistivity of this system is linear in temperature down to at least $1$ K, since $D$ is (almost) independent of temperature. 

\begin{figure}[h]
    \centering
    \includegraphics[width=1\linewidth]{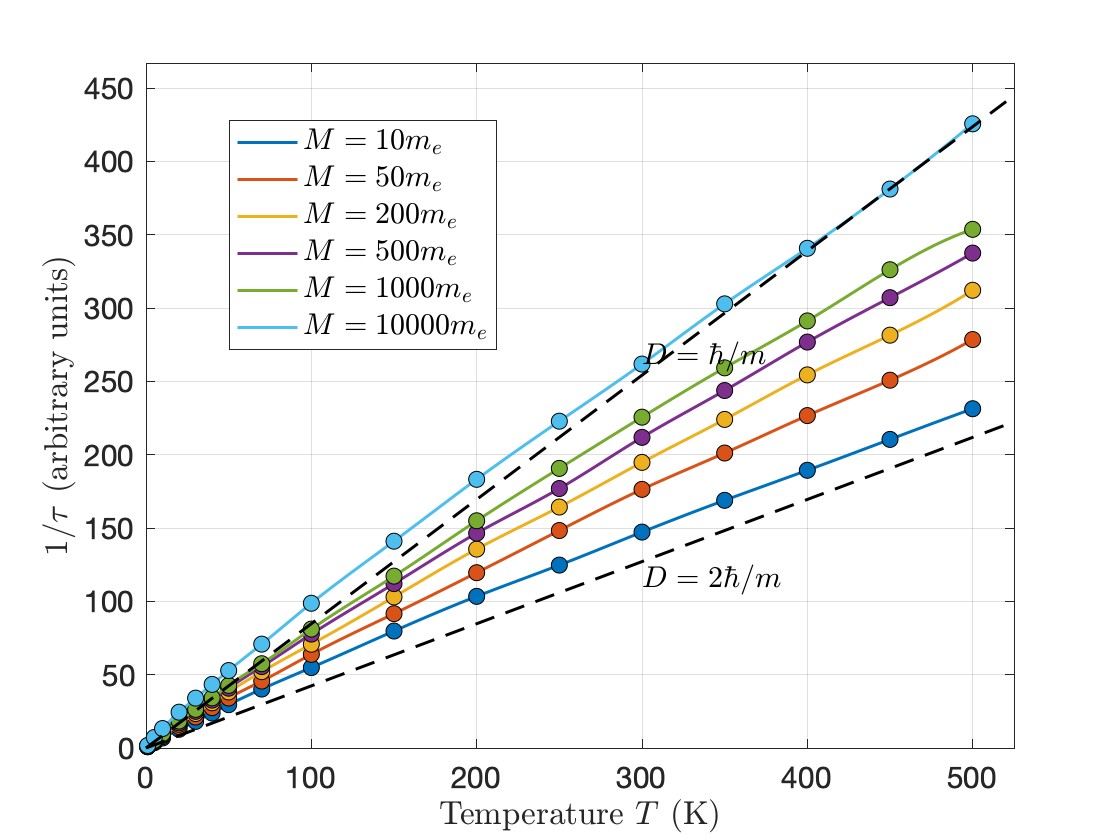}
    \caption{Linear Resistivity down to 1 K. Calculation based on Einstein relation and diffusion derived from the thermal Boltzmann distributed velocities of random impurities  model.}
    \label{T-linear}
\end{figure}

In Figure~\ref{T-linear}, we show the linear-in-temperature behavior, with different lines representing different impurity properties. 

\section{Phenomenological model}

 We now describe a nonthermal model, as a realization of Thouless scaling theory~\cite{abrahams1979scaling, edwards1972numerical}, that obeys Planckian diffusion $D \approx \hbar/m$.
 
 We seek to understand how Planckian diffusion emerges, using this model, which has chambers of area $A$ defined by irregular walls. The inspiration is a based on what we see happening in our quantum acoustic simulations, exhibiting transient localization~\cite{aydin_quantum_2024}. 
 
 

 Quantum mechanics controls the dwell time in the chambers. We ignore interference between different paths from chamber $n$ to chamber $m$. Although such interference occurs, the statistical average over realizations of the model, over all constructive and destructive interference events, would give a classical result and a 2D random walk. The quantum physics lies in the escape mechanism from the chambers.

We now seek the diffusion constant $D$ of this model.  For a 2D box of area $A$, with no internal features rising above the energy, the density of states is  $$\rho = 2 \pi m A/h^2.$$  A single channel ``leak'' allowing escape from a chaotic box gives a decay lifetime that broadens the levels by one level spacing, or $$\tau=\hbar \rho,$$ or $\tau = m A/h.$ So, if escape from one chamber to the next leads to a 2D random walk, with stepsize $d=\sqrt A$, and time between steps $\tau = m A/h,$ the 2D random walk diffusion constant is  $$D = d^2/4\tau =  2\pi \hbar/4m \sim \hbar/m,$$ independent of the confinement area $A.$ 


\begin{figure}[h]
    \centering
    \includegraphics[width=0.8\linewidth]{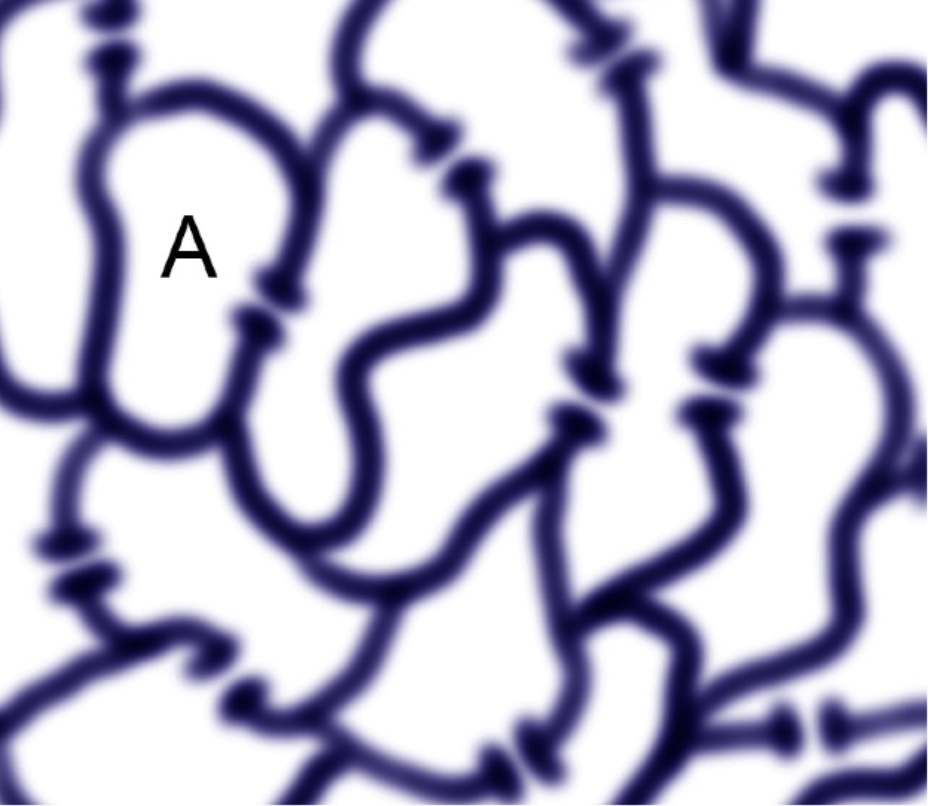}
    \caption{A model presented as a realization of a Thouless regime. Chambers of area roughly A are connected by single mode quantum channels or contacts; these may slowly open or close between chambers. This gives rise to Planckian, quantum diffusion, with $D\sim \hbar/m$, independent of A. There is no defined temperature, and indeed   Planckian diffusion  does not depend on there being a  temperature.  For thermal  systems,   In a thermal system and with an Einstein relation, Planckian diffusion does imply the Planckian speed limit, $\tau = \hbar/k_B T$. }
    \label{fig:enter-label}
\end{figure}

\section{Supporting cases}

\subsection{ Electrons on a solid hydrogen surface} There is a seemingly quite different experiment that is in astonishingly close analogy to the present problem, highlighting the universality of Planckian diffusion.  It will point the way to linear resistivity down to 0 Kelvin, without invoking an ephemeral electron fluid. 

\begin{figure}[h]
    \centering
\includegraphics[width=0.45\textwidth]{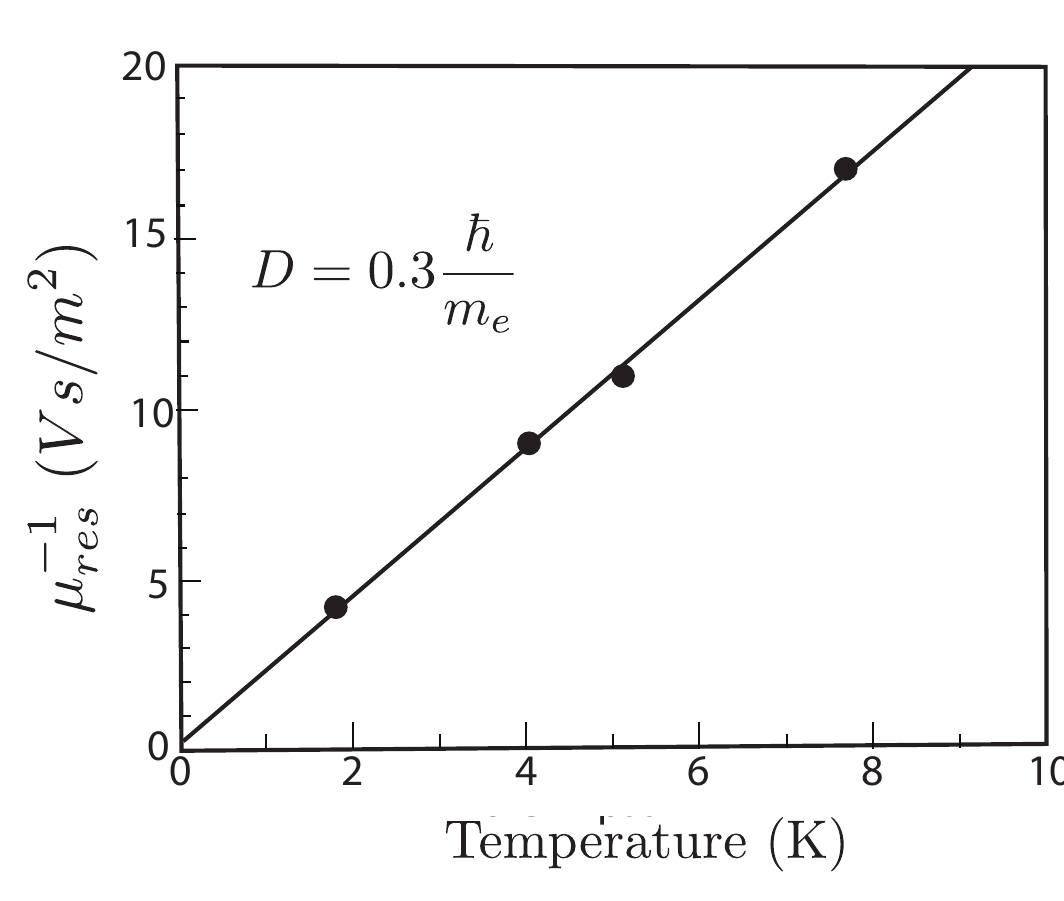}
    \caption{A re-drawn plot of the inverse residual mobility as a function of temperature, from reference \cite{adams_conductivity_1992}. The solid line is the best linear fit to the data. The slope was  interpreted by the author as  $D = 0.3\ \hbar/m_e$.  }
    \label{hdata}
\end{figure}

The new context, actually a well established experiment,  is the mobility of nondegenerate electrons on a 2D solid hydrogen surface, perturbed by  adsorbed Helium atoms, in work by Adams, and Adams and Paalanen\cite{adams_localization_1987,adams_conductivity_1992}. Planckian diffusion $D\sim\hbar/m_e$ emerges. The mobility is measured by magnetoresistance in a Corbino disk geometry (Figure~\ref{hdata}).

The physical situation is the following: Electrons are placed on the hydrogen solid surface; they don't penetrate far, and are trapped by roughness and defects.   Helium is then deposited on the surface. The gas is then pumped out and it takes some hours for the remaining Helium to come off the surface. While it is there, the electrons   diffuse and a mobility is established.  Our interpretation of this effect is the breaking of  localization by mobile Helium atoms, which here play the role of the thermal time evolution of the deformation potential. Thus the criteria for nondegenerate universal Planckian diffusion  under a temperature independent diffusion constant  $D\sim \hbar/m_e$ are met, (The effective mass is very close to the bare electron mass.)  That leads to linear resistivity with temperature, here in the 1-10 K region as seen in Figure~\ref{hdata}.

Figure \ref{hdata} shows that as long as localization is foiled by a moving perturbation, universal diffusion and linear resistivity   results, down to about 1 K in that figure.


\subsection{Lattice wave on quantum wave}
\begin{figure}[h]
    \centering
\includegraphics[width=0.35\textwidth]{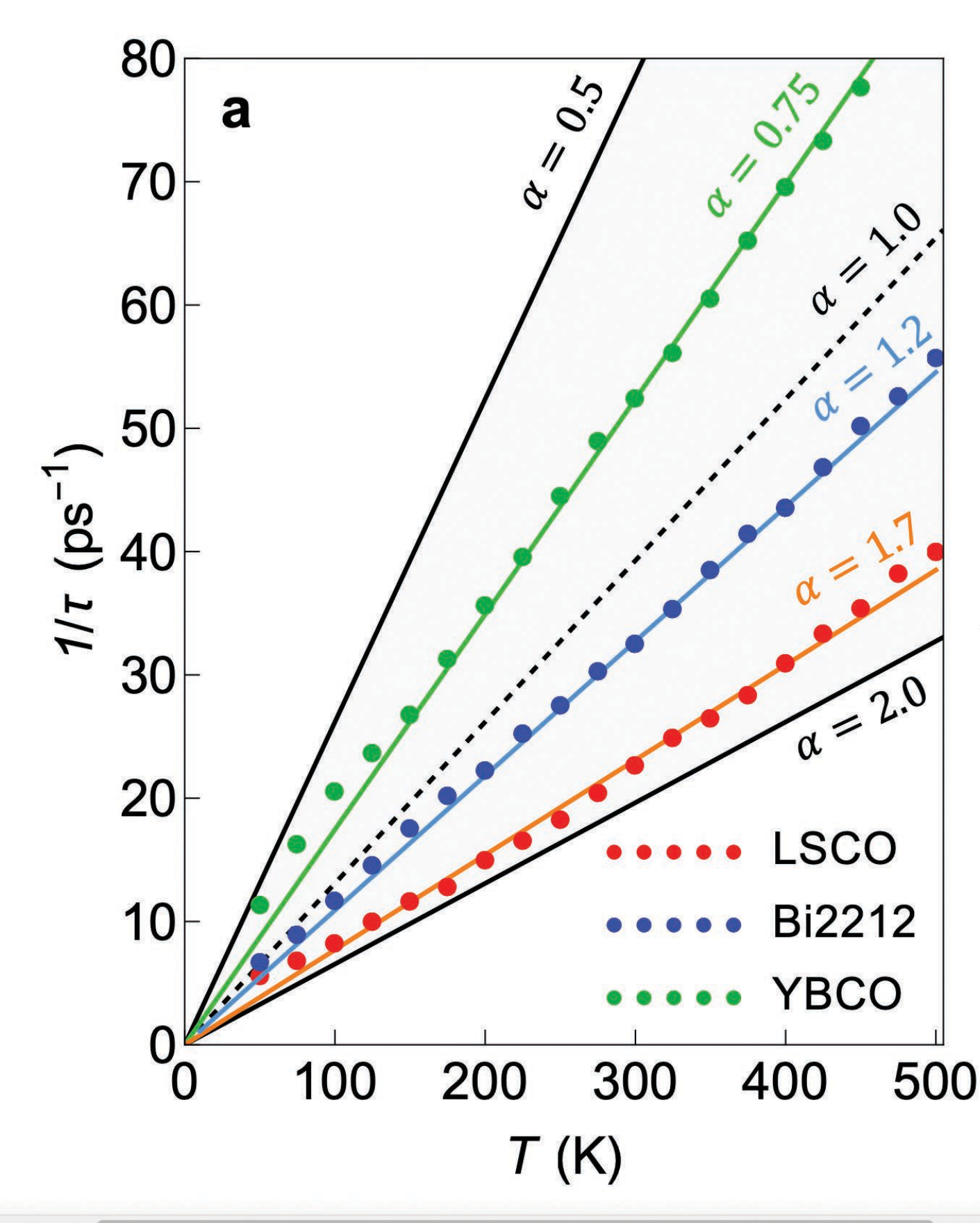}
    \caption{ For parameters appropriate to  three stange metals, the inverse relaxation time implied by our wave-on-wave simulations is givn as a function of temperature.}
    \label{planckian}
\end{figure}
In Figure~\ref{planckian} we cite  earlier results reinforcing the conclusions drawn here,   that linear resistivity at the correct Planckian slope prevails in the three strange metals investigated~\cite{aydin_quantum_2024}. The diffusion constant $D$ in each of the three cases was close to $\hbar/m^*$.  Each of the potentials, if frozen, would have localized the electron. This provides another example of broken Anderson localization leading to Planckian diffusion.

\section{Discussion}

 That Panckian diffusion is the natural outcome in {\it dynamical} disordered systems, insensitive to the specific details of the nature or the motion of the impurities, is demonstrated with numerical studies, a Thouless-like model, and an earlier experiment on solid hydrogen. 

All the four cases we reported above indicate that different combinations of localizing environments together with different  disturbances breaking the localization all suggest a universal outcome, with the same diffusion constant $D=\alpha \hbar/m$. The diffusion is insensitive to the localization length in the frozen limit, insensitive to the coupling of the carrier to the medium, and insensitive to the  temperature or even the existence of a temperature.
The phenomenon may be a universal consequence of the failed attempt to Anderson-localize. Many kinds  of disturbances can lead to its demise. The two behaviors are intimately linked.

We also identify mechanisms to depart from the 'universal' regime by moving the medium at either very low or very high velocities.  The localization length undoubtedly plays a critical role in defining what constitutes 'extreme,' with larger localization lengths transitioning to universality more readily under slow medium motion. If the potential evolves very slowly, localized energies of nearby bound states will exhibit avoided crossings. In the adiabatic regime, this results in predictable hopping of occupation between sites, with the hopping rate proportional to the rate of potential change, rendering it non-universal. This behavior does not violate Planckian diffusion but instead represents a distinct adiabatic 'sticky' phase. As the medium's motion accelerates, the system transitions rapidly to the diabatic 'slippery' phase, where Planckian diffusion dominates.



We see from our simulation, as the velocity of the medium increases from zero to a small fraction of the sound speed, the diffusion coefficient rises very rapidly but smoothly from zero to the Planckian rate. In the very slow, adiabatic limit, the coefficient does not have a well-defined lower bound as commonly assumed~\cite{hartnoll_colloquium_2022}.  This  supports a recent counterexample to the conjectured Planckian bound on transport ~\cite{poniatowski2021counterexample}, showing that the diffusion coefficient is possible to go below the Planckian limit.


It should be quite possible to design furthers tests  of this universality.  For example, the localization beautifully measured in reference~\cite{gavish_matter-wave_2005} could presumably be deliberately disturbed and the diffusion constant measured.

\section{Conclusion}

We have presented a new model of moving impurities clearly showing Planckian diffusion emerging from Anderson localization in two spatial dimensions.  We have also connected the results with a generic model of morphing weakly coupled chambers that would be Anderson localized if fixed, but yields Planckian diffusion when slowly morphing.  Referring to recent work We have shown the connections to the existing  examples of Planckian diffusion arising from the thermal deformation potential using strange metal parameters, and experiments involving electrons on solid hydrogen. This latter system shows ghost Planckian diffusion down to at least 2 Kelvin when the electrons are released from bondage by mobile Helium atoms on the surface. 

We anticipate that investigating the breakdown of Anderson localization via medium motion, both theoretically and experimentally, offers new pathways to understand its microscopic foundations.

Beyond this, we have find a new form of universality: the 'ghost' of a fundamental phenomenon—Planckian diffusion, which echoes Anderson localization. This concept not only encompasses and extends the Planckian speed limit but also matches the broad applicability of Anderson localization, broadening its scope to include moving media.

\section{Aknowledgements}

This work was supported by the U.S. Department of Energy under Grant No. DE-SC0025489. A.A. acknowledges financial support from the Sabanci University President’s Research Grant with project code F.A.CF.24-02932. A.M.G. thanks the Studienstiftung des Deutschen Volkes for financial support. J.K.-R. thanks the Oskar Huttunen Foundation for the financial support. 

\nocite{*}
\bibliography{refs}

\begin{thebibliography}{10}

\bibitem{anderson1958absence}
Philip~W Anderson.
\newblock Absence of diffusion in certain random lattices.
\newblock {\em Physical review}, 109(5):1492, 1958.

\bibitem{kramer1993localization}
Bernhard Kramer and Angus MacKinnon.
\newblock Localization: theory and experiment.
\newblock {\em Reports on Progress in Physics}, 56(12):1469, 1993.

\bibitem{lagendijk2009fifty}
Ad~Lagendijk, Bart~van Tiggelen, and Diederik~S Wiersma.
\newblock Fifty years of anderson localization.
\newblock {\em Physics today}, 62(8):24--29, 2009.

\bibitem{roati2008anderson}
Giacomo Roati, Chiara D’Errico, Leonardo Fallani, Marco Fattori, Chiara Fort, Matteo Zaccanti, Giovanni Modugno, Michele Modugno, and Massimo Inguscio.
\newblock Anderson localization of a non-interacting bose--einstein condensate.
\newblock {\em Nature}, 453(7197):895--898, 2008.

\bibitem{schwartz2007transport}
Tal Schwartz, Guy Bartal, Shmuel Fishman, and Mordechai Segev.
\newblock Transport and anderson localization in disordered two-dimensional photonic lattices.
\newblock {\em Nature}, 446(7131):52--55, 2007.

\bibitem{hu2008localization}
Hefei Hu, A~Strybulevych, JH~Page, Sergey~E Skipetrov, and Bart~A van Tiggelen.
\newblock Localization of ultrasound in a three-dimensional elastic network.
\newblock {\em Nature Physics}, 4(12):945--948, 2008.

\bibitem{billy2008direct}
Juliette Billy, Vincent Josse, Zhanchun Zuo, Alain Bernard, Ben Hambrecht, Pierre Lugan, David Cl{\'e}ment, Laurent Sanchez-Palencia, Philippe Bouyer, and Alain Aspect.
\newblock Direct observation of anderson localization of matter waves in a controlled disorder.
\newblock {\em Nature}, 453(7197):891--894, 2008.

\bibitem{white_observation_2020}
Donald~H. White, Thomas~A. Haase, Dylan~J. Brown, Maarten~D. Hoogerland, Mojdeh~S. Najafabadi, John~L. Helm, Christopher Gies, Daniel Schumayer, and David A.~W. Hutchinson.
\newblock Observation of two-dimensional {Anderson} localisation of ultracold atoms.
\newblock {\em Nature Communications}, 11(1):4942, October 2020.

\bibitem{kondov2011three}
SS~Kondov, WR~McGehee, JJ~Zirbel, and B~DeMarco.
\newblock Three-dimensional anderson localization of ultracold matter.
\newblock {\em Science}, 334(6052):66--68, 2011.

\bibitem{jendrzejewski2012three}
Fred Jendrzejewski, Alain Bernard, Killian Mueller, Patrick Cheinet, Vincent Josse, Marie Piraud, Luca Pezz{\'e}, Laurent Sanchez-Palencia, Alain Aspect, and Philippe Bouyer.
\newblock Three-dimensional localization of ultracold atoms in an optical disordered potential.
\newblock {\em Nature Physics}, 8(5):398--403, 2012.

\bibitem{afmTL}
Simone Fratini, Didier Mayou, and Sergio Ciuchi.
\newblock The transient localization scenario for charge transport in crystalline organic materials.
\newblock {\em Advanced Functional Materials}, 26(14):2292--2315, 2016.

\bibitem{heller22}
Donghwan Kim, Alhun Aydin, Alvar Daza, Kobra~N. Avanaki, Joonas Keski-Rahkonen, and Eric~J. Heller.
\newblock Coherent charge carrier dynamics in the presence of thermal lattice vibrations.
\newblock {\em Phys. Rev. B}, 106:054311, Aug 2022.

\bibitem{zimmermann_rise_2024}
Yoel Zimmermann, Joonas Keski-Rahkonen, Anton~M. Graf, and Eric~J. Heller.
\newblock Rise and {Fall} of {Anderson} {Localization} by {Lattice} {Vibrations}: {A} {Time}-{Dependent} {Machine} {Learning} {Approach}.
\newblock {\em Entropy}, 26(7):552, June 2024.

\bibitem{w}
Joonas Keski-Rahkonen, Xiaoyu Ouyang, Shaobing Yuan, Anton~M. Graf, Alhun Aydin, and Eric~J. Heller.
\newblock Quantum-acoustical drude peak shift.
\newblock {\em Phys. Rev. Lett.}, 132:186303, May 2024.

\bibitem{Aydin24}
Alhun Aydin, Joonas Keski-Rahkonen, and Eric~J. Heller.
\newblock Quantum acoustics unravels planckian resistivity.
\newblock {\em Proceedings of the National Academy of Sciences}, 121(28):e2404853121, 2024.

\bibitem{aydin_quantum_2024}
Alhun Aydin, Joonas Keski-Rahkonen, and Eric~J. Heller.
\newblock Quantum acoustics unravels {Planckian} resistivity.
\newblock {\em Proceedings of the National Academy of Sciences}, 121(28):e2404853121, July 2024.
\newblock arXiv:2303.06077 [cond-mat].

\bibitem{chung2021ultra}
Yoon~Jang Chung, KA~Villegas~Rosales, KW~Baldwin, PT~Madathil, KW~West, M~Shayegan, and LN~Pfeiffer.
\newblock Ultra-high-quality two-dimensional electron systems.
\newblock {\em Nature Materials}, 20(5):632--637, 2021.

\bibitem{hwang2013electronic}
EH~Hwang and S~Das~Sarma.
\newblock Electronic transport in two-dimensional si: P $\delta$-doped layers.
\newblock {\em Physical Review B—Condensed Matter and Materials Physics}, 87(12):125411, 2013.

\bibitem{wetzelaer_validity_2011}
G.~A.~H. Wetzelaer, L.~J.~A. Koster, and P.~W.~M. Blom.
\newblock Validity of the {Einstein} {Relation} in {Disordered} {Organic} {Semiconductors}.
\newblock {\em Physical Review Letters}, 107(6):066605, August 2011.

\bibitem{abrahams1979scaling}
Elihu Abrahams, Philip~W Anderson, Donald~C Licciardello, and Tiruppattur~V Ramakrishnan.
\newblock Scaling theory of localization: Absence of quantum diffusion in two dimensions.
\newblock {\em Physical Review Letters}, 42(10):673, 1979.

\bibitem{edwards1972numerical}
JT~Edwards and DJ~Thouless.
\newblock Numerical studies of localization in disordered systems.
\newblock {\em Journal of Physics C: Solid State Physics}, 5(8):807, 1972.

\bibitem{adams_localization_1987}
Philip~W. Adams and Mikko~A. Paalanen.
\newblock Localization in a {Nondegenerate} {Two}-{Dimensional} {Electron} {Gas}.
\newblock {\em Physical Review Letters}, 58(20):2106--2109, May 1987.

\bibitem{adams_conductivity_1992}
P.W. Adams.
\newblock The conductivity and mobility of {2D} nondegenerate electrons in the strong localization regime.
\newblock {\em Surface Science}, 263(1-3):663--667, February 1992.

\bibitem{hartnoll_colloquium_2022}
Sean~A. Hartnoll and Andrew~P. Mackenzie.
\newblock \textit{{Colloquium}} : {Planckian} dissipation in metals.
\newblock {\em Reviews of Modern Physics}, 94(4):041002, November 2022.

\bibitem{poniatowski2021counterexample}
Nicholas~R Poniatowski, Tarapada Sarkar, Ricardo~PSM Lobo, Sankar Das~Sarma, and Richard~L Greene.
\newblock Counterexample to the conjectured planckian bound on transport.
\newblock {\em Physical Review B}, 104(23):235138, 2021.

\bibitem{gavish_matter-wave_2005}
Uri Gavish and Yvan Castin.
\newblock Matter-{Wave} {Localization} in {Disordered} {Cold} {Atom} {Lattices}.
\newblock {\em Physical Review Letters}, 95(2):020401, July 2005.

\bibitem{ahn2022density}
Seongjin Ahn and Sankar Das~Sarma.
\newblock Density-dependent two-dimensional optimal mobility in ultra-high-quality semiconductor quantum wells.
\newblock {\em Physical Review Materials}, 6(1):014603, 2022.

\bibitem{gosling2021universal}
Jonathan~H Gosling, Oleg Makarovsky, Feiran Wang, Nathan~D Cottam, Mark~T Greenaway, Amalia Patan{\`e}, Ricky~D Wildman, Christopher~J Tuck, Lyudmila Turyanska, and T~Mark Fromhold.
\newblock Universal mobility characteristics of graphene originating from charge scattering by ionised impurities.
\newblock {\em Communications Physics}, 4(1):30, 2021.

\bibitem{wolf1985weak}
Pierre-Etienne Wolf and Georg Maret.
\newblock Weak localization and coherent backscattering of photons in disordered media.
\newblock {\em Physical review letters}, 55(24):2696, 1985.

\bibitem{grissonnanche_linear-temperature_2021}
Gaël Grissonnanche, Yawen Fang, Anaëlle Legros, Simon Verret, Francis Laliberté, Clément Collignon, Jianshi Zhou, David Graf, Paul~A. Goddard, Louis Taillefer, and B.~J. Ramshaw.
\newblock Linear-in temperature resistivity from an isotropic {Planckian} scattering rate.
\newblock {\em Nature}, 595(7869):667--672, July 2021.

\bibitem{bruin_similarity_2013}
J.~A.~N. Bruin, H.~Sakai, R.~S. Perry, and A.~P. Mackenzie.
\newblock Similarity of {Scattering} {Rates} in {Metals} {Showing} \textit{{T}} -{Linear} {Resistivity}.
\newblock {\em Science}, 339(6121):804--807, February 2013.

\bibitem{legros_universal_2019}
A.~Legros, S.~Benhabib, W.~Tabis, F.~Laliberté, M.~Dion, M.~Lizaire, B.~Vignolle, D.~Vignolles, H.~Raffy, Z.~Z. Li, P.~Auban-Senzier, N.~Doiron-Leyraud, P.~Fournier, D.~Colson, L.~Taillefer, and C.~Proust.
\newblock Universal {T}-linear resistivity and {Planckian} dissipation in overdoped cuprates.
\newblock {\em Nature Physics}, 15(2):142--147, February 2019.

\bibitem{yang_signatures_2022}
Chao Yang, Haiwen Liu, Yi~Liu, Jiandong Wang, Dong Qiu, Sishuang Wang, Yang Wang, Qianmei He, Xiuli Li, Peng Li, Yue Tang, Jian Wang, X.~C. Xie, James~M. Valles, Jie Xiong, and Yanrong Li.
\newblock Signatures of a strange metal in a bosonic system.
\newblock {\em Nature}, 601(7892):205--210, January 2022.

\bibitem{morong_simulation_2015}
W.~Morong and B.~DeMarco.
\newblock Simulation of {Anderson} localization in two-dimensional ultracold gases for pointlike disorder.
\newblock {\em Physical Review A}, 92(2):023625, August 2015.

\bibitem{chakrabarty_quantum_2023}
Saurish Chakrabarty and Zohar Nussinov.
\newblock Quantum equilibration and measurements -- bounds on speeds, {Lyapunov} exponents, and transport coefficients obtained from the uncertainty relations and their comparison with experimental data, February 2023.
\newblock arXiv:2303.00021 [cond-mat].

\bibitem{luciuk_observation_2017}
C.~Luciuk, S.~Smale, F.~Böttcher, H.~Sharum, B.A. Olsen, S.~Trotzky, T.~Enss, and J.H. Thywissen.
\newblock Observation of {Quantum}-{Limited} {Spin} {Transport} in {Strongly} {Interacting} {Two}-{Dimensional} {Fermi} {Gases}.
\newblock {\em Physical Review Letters}, 118(13):130405, March 2017.

\bibitem{hartman_upper_2017}
Thomas Hartman, Sean~A. Hartnoll, and Raghu Mahajan.
\newblock Upper {Bound} on {Diffusivity}.
\newblock {\em Physical Review Letters}, 119(14):141601, October 2017.

\bibitem{hu_universal_2017}
Tao Hu, Yinshang Liu, Hong Xiao, Gang Mu, and Yi-feng Yang.
\newblock Universal linear-temperature resistivity: possible quantum diffusion transport in strongly correlated superconductors.
\newblock {\em Scientific Reports}, 7(1):9469, August 2017.

\bibitem{hartnoll_theory_2015}
Sean~A. Hartnoll.
\newblock Theory of universal incoherent metallic transport.
\newblock {\em Nature Physics}, 11(1):54--61, January 2015.

\bibitem{lanzara_evidence_2001}
A.~Lanzara, P.~V. Bogdanov, X.~J. Zhou, S.~A. Kellar, D.~L. Feng, E.~D. Lu, T.~Yoshida, H.~Eisaki, A.~Fujimori, K.~Kishio, J.-I. Shimoyama, T.~Noda, S.~Uchida, Z.~Hussain, and Z.-X. Shen.
\newblock Evidence for ubiquitous strong electron–phonon coupling in high-temperature superconductors.
\newblock {\em Nature}, 412(6846):510--514, August 2001.

\bibitem{zaanen_planckian_2019}
Jan Zaanen.
\newblock Planckian dissipation, minimal viscosity and the transport in cuprate strange metals.
\newblock {\em SciPost Physics}, 6(5):061, May 2019.

\bibitem{stephen_weak_1987}
Michael~J. Stephen.
\newblock Weak localization and the conductivity of nondegenerate electrons.
\newblock {\em Physical Review B}, 36(10):5663--5664, October 1987.

\bibitem{kovtun_viscosity_2005}
P.~K. Kovtun, D.~T. Son, and A.~O. Starinets.
\newblock Viscosity in {Strongly} {Interacting} {Quantum} {Field} {Theories} from {Black} {Hole} {Physics}.
\newblock {\em Physical Review Letters}, 94(11):111601, March 2005.

\end{thebibliography}
\bibliographystyle{unsrt}
\end{document}